\documentclass[prb,floatfix,twocolumn,secnumarabic,amssymb, amsmath,nobibnotes,aps,prl,superscriptaddress,longbibliography]{revtex4-1}

\usepackage{graphicx}% Include figure files
\usepackage{bm}% bold math
\usepackage{makecell} % multi-line text in tables
\usepackage{color}
\usepackage{gensymb}
\usepackage{lineno}

\begin{document}

\title{Bayesian Inference of Atomistic Structure in Functional Materials} %prediction}

\author{Milica Todorovi\'{c}}%
\email[Corresponding author:\\]{milica.todorovic@aalto.fi}
\affiliation{Department of Applied Physics, Aalto University, P.O. Box 11100, Aalto FI-00076, Finland}%

\author{Michael U. Gutmann} 
\affiliation{School of Informatics, University of Edinburgh, 10 Crichton Street, Edinburgh, EH8 9AB, United Kingdom}

\author{Jukka Corander}
\affiliation{Institute of Basic Medical Sciences, University of Oslo, Sognsvannsveien 9, 0372 Oslo, Norway}
\affiliation{Department of Mathematics and Statistics, University of Helsinki, P.O. Box 68, Helsinki, FI-00014, Finland}

\author{Patrick Rinke}%
\affiliation{Department of Applied Physics, Aalto University, P.O. Box 11100, Aalto FI-00076, Finland}%

\date{\today}

%\linenumbers
\begin{abstract}
\noindent Tailoring the functional properties of advanced organic/inorganic heterogeneous devices to their intended technological applications requires knowledge and control of the microscopic structure inside the device. Atomistic quantum mechanical simulation methods deliver accurate energies and properties for individual configurations, however, finding the most favourable configurations remains computationally prohibitive. We propose a 'building block'-based Bayesian Optimization Structure Search (BOSS) approach for addressing extended organic/inorganic interface problems and demonstrate its feasibility in a molecular surface adsorption study. In BOSS, a Bayesian model identifies material energy landscapes in an accelerated fashion from atomistic configurations sampled during active learning. This allowed us to identify several most favorable molecular adsorption configurations for $\mathrm{\textbf{C}}_{\textbf{60}}$ on the (101) surface of $\mathrm{\textbf{TiO}}_{\mathbf{2}}$ anatase and clarify the key molecule-surface interactions governing structural assembly. Inferred structures were in good agreement with detailed experimental images of this surface adsorbate, demonstrating good predictive power of BOSS and opening the route towards large-scale surface adsorption studies of molecular aggregates and films.  
\end{abstract}

\pacs{Valid PACS appear here}% PACS, the Physics and Astronomy
\maketitle

Frontier technologies are increasingly based on functional hybrid materials - engineered blends of organic molecules and inorganic crystals that harness and enhance the functional properties of both substances to perform specific tasks. Organic/inorganic heterostructures and metal-organic frameworks are key components for smart sensors, membranes and coatings, novel optoelectronic and fuel cell technologies, with further applications in data storage, quantum engineering and nanophotonics on the horizon \cite{Theobald/etal:2003,Barth:2005dl,Grill/etal:2007,Schlesinger/etal:2015,DennyJr.2016,Huang2016,Geng/etal:2017,Song/etal:2017}. Despite outstanding component materials, engineering the microscopic structure of complex heterostructures to tailor their properties towards desired functionality remains a fundamental challenge in physics, chemistry and materials science. It means bypassing the pitfalls of interface artifacts, defects and unfavourable self-assembled structures that degrade overall device performance.

Understanding the microscopic structural details of advanced organic/inorganic material blends has emerged as the primary route towards controlling and engineering the functionality of hybrid materials \cite{Barth:2005dl,Howarth2016}.  Here computational studies lead the way \cite{vonLilienfeld:2014fz,2013NatMa..12..191C}, since nanoscale experimental measurement techniques frequently lack the necessary atomistic detail, and traditional trial-and-error tests are costly and time-consuming. \emph{Ab initio} methods like density functional theory (DFT) are especially predictive in simulations of hybrid materials because they accurately describe the delicate interplay of microscopic interactions 
%%% FIG 1. %%%
\begin{figure}[tbh]
\centering
\includegraphics[width=.8\linewidth]{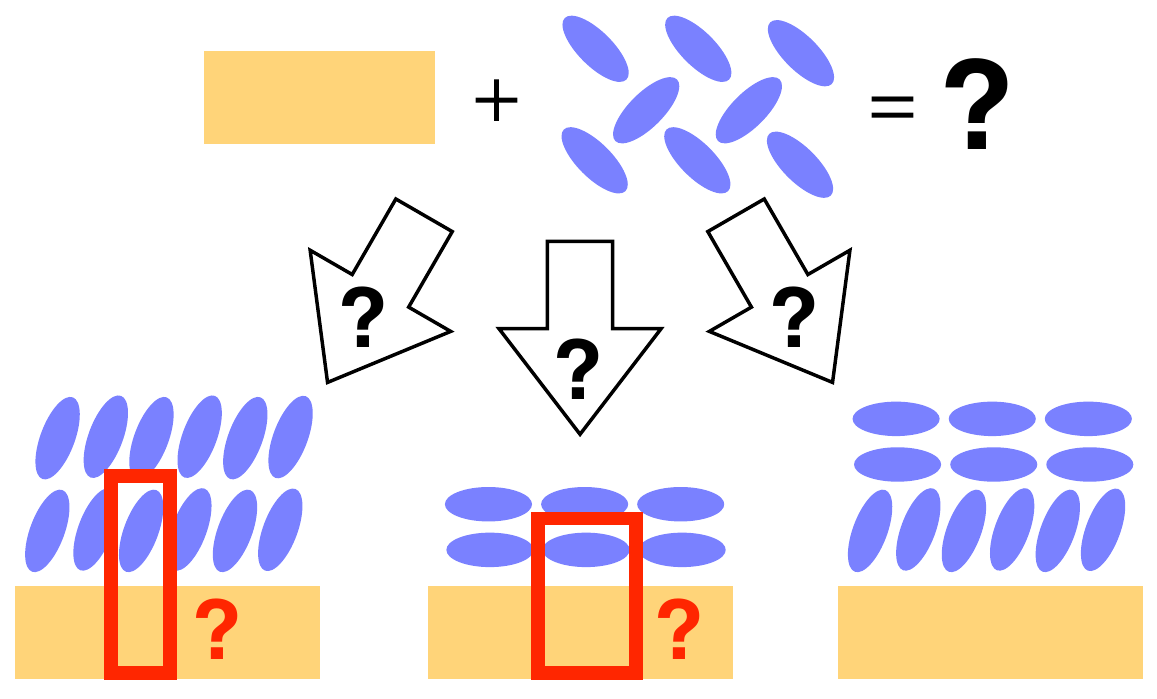}
\caption{Inside devices, various thin-film morphologies composed from organic molecules (blue oval) may be formed at the interface with a crystalline solid (yellow). In the first step towards a large-scale Bayesian structure search of monolayer morphologies, we focus on inferring the configuration of a single molecule adsorbate (shown in red).}
\label{fig1}
\end{figure}
%%%%%%%%%%%%%%
(e.g. electrostatics, dispersion, bond formation and charge transfer) that direct structural assembly \cite{Jones:2015wt}. DFT maps the atomic structure of a material onto an intrinsic energy, with lower energies indicating more stable material polymorphs. Theoretical structure prediction methods focus on exploring the resulting configurational phase-space, the potential energy surface (PES) \cite{1983Sci...220..671K,2005PhRvL..95e5501G}. Extensive PES sampling by DFT is computationally prohibitive and intractable. In practice it must be reduced to comparing several ‘most-likely’ structures, which is unreliable in complex materials. 

For this reason, hybrid organic/inorganic interfaces present a special challenge for structure search methods. As illustrated in Fig. \ref{fig1}, their PES is complicated by the variety of different morphologies that molecular films can adopt against the solid material. Moreover, the large size of functional molecules means that extensive simulation cells (large lengthscales) are needed to describe molecular film morphologies, making computations particularly expensive. 
%%% FIG 2. %%%
\begin{figure*}[t]
\centering
\includegraphics[width=16cm]{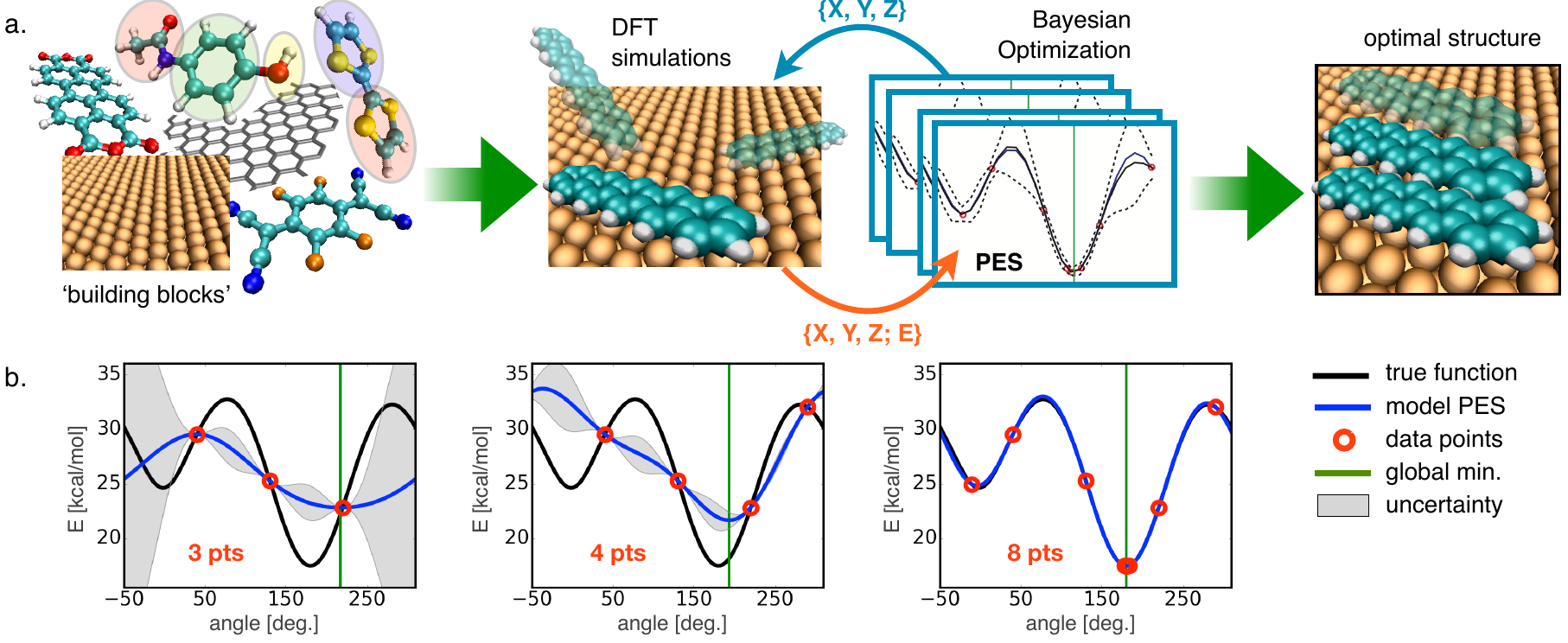}
\caption{Illustration of a typical BOSS application. a. Schematic of key steps in BOSS structure search at the inorganic surface: from the choice of materials and building blocks, through selection of the BO degrees of freedom and the iterative optimisation, towards the inferred individual adsorbate and thin-film structures; b. Example of BOSS iterative inference of a simple 1-dimensional (1D) PES featuring a global and local minimum. The GP native uncertainty (gray areas) facilitates exploratory data sampling. The global minimum location and the entire PES are learned in eight data acquisitions.}
\label{fig2}
\end{figure*}
%%%%%%%%%%%%%%
% ML solutions

To address this structure search problem, we harness the power of AI methods.
Recently, AI and machine learning (ML) algorithms were coupled with DFT to approximate the PES \cite{Behler:2014jc,Bartok:2010fj,2015PhRvL.114i6405L} or improve sampling and accelerate structure prediction in single material clusters and solids 
\cite{2008PhRvB..78f4102D,Wang:2012dn,2013PhRvB..87c5125N,2013PhRvL.111m5501B,Kiyohara:2016hy,2016NatCo...711241X}. Their application to heterostructures is not straightforward, and they may not scale up to required sizes. In some cases, framework setup and the choice of ML parameters was found to affect the results \cite{Behler:2014jc,2014NJPh...16l3016B}. 
Many schemes rely on large data sets with 1,000-10,000 sampled points \cite{Ueno:2016kj}, which are costly to compute. Our ideal method would need to be (i) \emph{efficient} (minimal sampling costs), (ii) \emph{accurate} (both in robust model convergence and DFT chemical accuracy), (iii) \emph{comprehensive} (delivering the entire PES information of global and local minima), (iv) \emph{transferable} (minimal dependence on ML parameters), (v) \emph{versatile} (adaptable to targeting properties, structural prescreening, etc.), (vi) \emph{flexible} (easily combined with other schemes) and (vii) truly \emph{multi-scale} in its scope.

% our solution
Here, we propose an AI-based structure search scheme that is capable of accelerated and unbiased PES computation, and can be extended to large length-scales while minimising the amount of configurational sampling. The Bayesian Optimisation Structure Search (BOSS) method, illustrated in Fig. \ref{fig2}a, couples state-of-the-art DFT or quantum chemistry treatment with the Bayesian Optimisation (BO) technique for complex optimisation tasks. 

\textbf{Active PES learning with BOSS}. Approximate Bayesian Computation \cite{doi:10.1093/sysbio/syw077} is a class of likelihood-free inference (LFI) methods where data sampling involves complex evaluation. It has recently been combined with BO \cite{Brochu:2010um} to accelerate model prediction where data evaluation is also costly. In this work, we adapted the resulting BOLFI scheme \cite{Gutmann:wf} to search for minima of the PES in an arbitrary phase space using Gaussian Process (GP) models. 
BOSS utilises an advanced DFT framework designed for efficient first principles materials simulations on supercomputer infrastructures \cite{2009CoPhC.180.2175B}. Each data point is a DFT total energy representing an atomistic configuration. 

BOSS employs GPs to fit a surrogate PES model to DFT data points, then refines it by acquiring more data points through a smart sampling strategy (see Fig.  \ref{fig2}b.). The most likely PES model for given data is the GP posterior mean, which can be traversed by minimisation algorithms to determine all minima and their locations in phase space. The GP posterior variance reflects the lack of confidence in the probabilistic PES model, which vanishes at the datapoints, and rises in unexplored areas of phase space. In analogy with the 1D example in Fig.  \ref{fig2}b., BOSS actively learns the every point of the PES in N dimensions and across the defined phase space until convergence is achieved.

Smart sampling of new configurations allows BOSS to make accurate DFT-based predictions despite the DFT's computational cost. Our chosen algorithm for sequential acquisition of new energy points combines \emph{exploration} (searching less visited areas) with \emph{exploitation} (searching low energy areas) to determine the PES global minimum with as few data points as possible. 
%This sampling procedure is illustrated in Fig. \ref{fig2}b. 
Such a strategy, encapsulated in the exploratory Lower Confidence Bound (eLCB) acquisition function \cite{Cox:1992kb, Brochu:2010um}, ensures fast determination of the global minimum. We employ an acquisition function that increasingly favours exploration with rising search dimensionality and iteration step \cite{Gutmann:wf}.

%%% FIG 3. %%%
\begin{figure*}[htb]
\centering
\includegraphics[width=16cm]{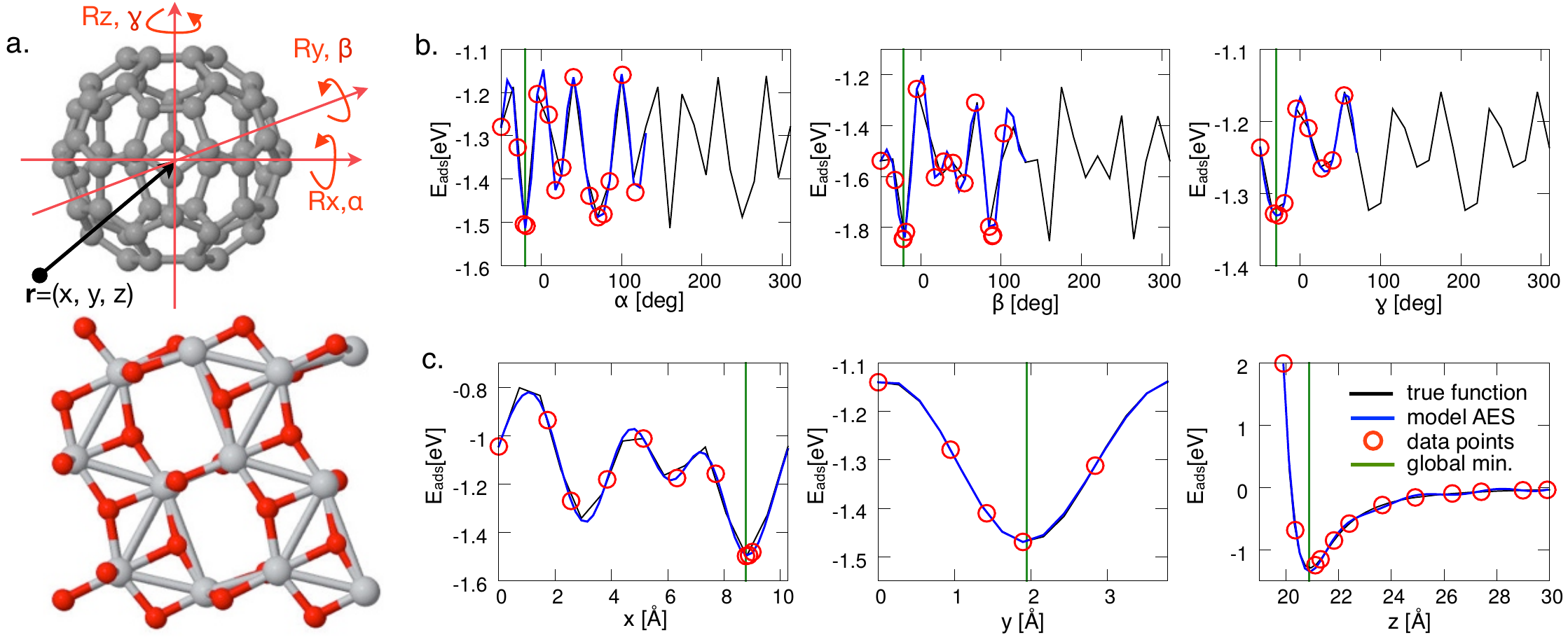}
\caption{BOSS applied to the $\mathrm{C}_{60}$/$\mathrm{TiO}_2$ adsorption problem. a. Atomistic model of $\mathrm{C}_{60}$ on the (101) surface $\mathrm{TiO}_2$ anatase in the reference configuration, with the energetically dominant degrees of freedom for the molecule indicated in black (translational motion) and red (rotational motion); b. Comparison of the converged 1D AES with the true function for all rotational variables; c. Comparison of the converged 1D AES with the true function for all translational degrees of freedom. Learning in b. and c. was initialised with 5 quasi-random points and the models converged in up to 7 BO acquisitions. 1D searches were carried out with the other variables fixed to reference values, as illustrated in a. and described in the Methods section.}\label{fig3}
\end{figure*}
%%%%%%%%%%%%%%

A common feature of structure search in complex heterogenous materials is the presence of rigid organic and inorganic structures, (aromatic rings, functional groups), where structure change is confined to small bond adjustments, without bond re-arrangements. To expedite structure search over large numbers of atoms, we follow other schemes \cite{2007JChPh.127a4705B,Oganov:wk} and  fix these internal components of the material to rigid 'building block’ components. Such an approach is suitable to describe molecular physisorption and some chemisorption via anchoring groups, both common at hybrid interfaces. The choice of building blocks is motivated by chemical rules, and expedites the search by confining it to configurational phase space, instead of full chemical phase space.

In the long-term, BOSS can be used to predict the structure of organic/inorganic interfaces by identifying the most stable organic thin film morpholgies on inorganic substrates. The procedure is illustrated in Fig. \ref{fig2}a: once the simulation 'building blocks' are identified, the learning would progress from single adsorbates to molecular aggregates and monolayers. While some methods acquire single adsorption configurations by intuition and focus on complex lattice-based film morphology search \cite{2017NanoL..17.4453O,2017NatCo...814463P}, we aim to treat both the molecular adsorbates and aggregates within the BOSS framework by increasing search degrees of freedom.

Learning the individual molecule-surface interactions and structure is a \emph{key step}, which is demonstrated here by applying BOSS on inferring the structure of a single molecular surface adsorbate.
In this manuscript, we conducted a structural study of a fullerene molecule on the (101) surface of $\mathrm{TiO}_2$ anatase. Both are functional materials frequently employed in organic optoelectronics \cite{Gratzel:1998io,Yoo:2007ju,Cheyns:2007gz}. 
To verify AI predictions, inferred structures were compared to the high-resolution atomic force miscroscopy (AFM) images \cite{2015NanoL..15.2257M} of this surface system.

\textbf{Results and Discussion.} The atomistic simulation model of $\mathrm{C}_{60}$ on the $\mathrm{TiO}_2$ anatase surface is presented in Fig. \ref{fig3}a. The surface slab and fullerene cage were defined as building blocks. Stable molecular adsorbate structures are the atomistic configurations that minimise the adsorption energy, so BOSS was set to learn the adsorption energy surface (AES). The adsorption energy depends on the molecular position above the surface, represented by the molecular centre of mass $r$=[$x$, $y$, $z$], and its orientation towards the surface. The latter was described by angles of rotation $\alpha$, $\beta$ and $\gamma$ with respect to Cartesian axes of rotation $\mathrm{R}_{x}$, $\mathrm{R}_{y}$ and $\mathrm{R}_{z}$, respectively.
%%% FIG 4. %%%
\begin{figure}[bht]
\centering
\includegraphics[width=0.9\linewidth]{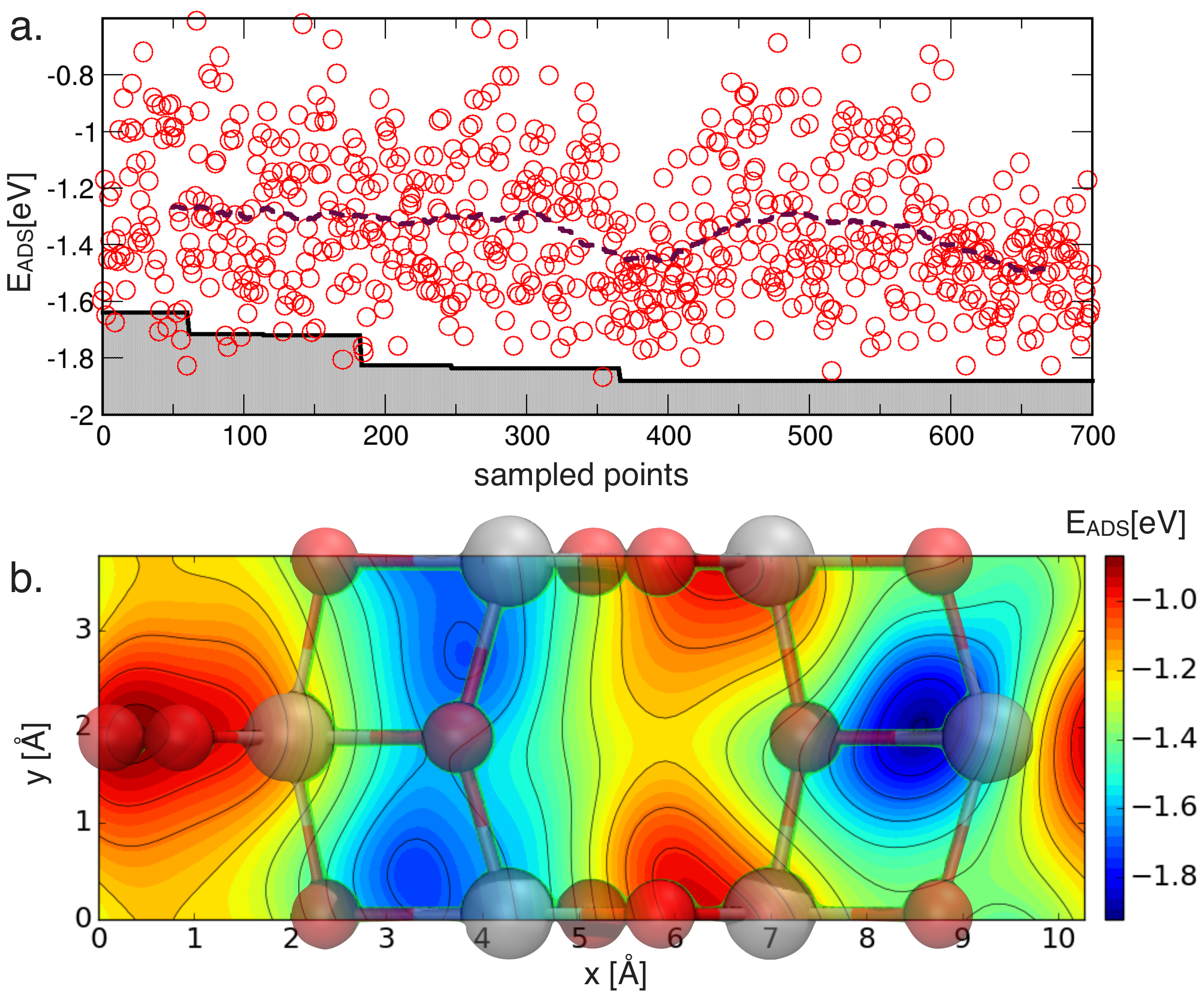}
\caption{5D BOSS search results. a. Convergence of the $\mathrm{E}_{\mathrm{ADS}}$ computed from the BOSS global minimum prediction during active learning (black line). The accuracy of the inferred result improved with strategic 5D configurational sampling (red data points, with running average shown in dashed line). b. 2D cross-section of the 5D BOSS search illustrating $x$-$y$ molecular translation, extracted at the 5D global minimum after 700 data acquisitions. The first $\mathrm{TiO}_2$  surface layer shown in overlay reveals the correlation between the global minimum (deep blue) and the $\mathrm{Ti}_{5c}$ surface site (gray atom).}
\label{fig4}
\end{figure}

The full AES is a 6-dimensional (6D) function of rotational and translational degrees of freedom $E_{\mathrm{AES}}$=E($\alpha$, $\beta$, $\gamma$, $x$, $y$, $z$). In Figs. \ref{fig3}b.\ and \ref{fig3}c.\ we present a BOSS investigation into each of these variables separately, which revealed the approximate AES variation from -1eV to -2eV. The $z$ variable was found to produce only a vertical shift in the adsorption energy. The location of the minima in other dimensions did not change with z, so we fixed it and carried out the full adsorption site BOSS search in 5D.

Fig. \ref{fig4}a. illustrates the refinement of the predicted 5D global minimum with iterative configurational sampling. The lowest observed adsorption energy $\mathrm{E}_{\mathrm{ADS}}$ (computed from BOSS-predicted global minimum locations) converged after 370 sampled configurations to a value of -1.88eV. Improvement of the global minimum prediction could be correlated to instances of visiting low energy configurations, chosen strategically from a vast 5D phase space. However, most model refinement proceeded with input from less relevant configurations, on average in the region 0.5eV above the predicted global minimum (after 400 iterations, the average acquisitions shifted to lower values, suggesting that the model is exploring near local minima). 
A physically meaningful 5D model of the  $\mathrm{E}_{\mathrm{ADS}}$  landscape (consistent with the symmetries of the DFT simulations) converged after 670 data acquisitions. Fig  \ref{fig4}b. shows the $x-y$ cross-section extracted from the 5D model at iteration 700.  The AES landscape correlates well with the two sloping terraces of the $\mathrm{TiO}_2$ surface.
BOSS typically finds the global minimum quickly, while more data is needed to refine the entire PES model.

\textbf{Chemical insight from AI models.} The chemically natural assignment of 'building blocks' means that resulting energy curves lend themselves readily to human interpretation. Already the preliminary 1D BOSS simulations revealed a wealth of information about the binding and structure at the $\mathrm{C}_{60}$/$\mathrm{TiO}_2$ interface. This helped us to determine the key binding sites on both the molecule and the surface.

Translations of the molecule across the surface produced slowly-varying energies with few minima (Fig. \ref{fig3}c.), closely reflecting anatase corrugation. The surface adsorption site was the $\mathrm{Ti}_{5c}$ or the $\mathrm{O}_{3c}$ one, depending on molecular orientation.
Molecular rotation gave rise to complex fast-varying AES curves with multiple deep minima (Fig. \ref{fig3}b.), as expected from the high symmetry of the $\mathrm{C}_{60}$ cage. 
By analyzing 1D global minima in $\beta$ (-1.85~eV) and $\alpha$ (-1.50~eV), we determined the active sites on the molecular cage to be the hexagonal facet and the $\mathrm{C}_{\mathrm{h}}$-$\mathrm{C}_{\mathrm{h}}$ bond between them (respectively). 

These findings are consistent with the global minimum structure inferred in the 5D AI search. Molecular rotation was the energetically dominant factor for surface adsorption.
The global minimum orientation of the physisorbed $\mathrm{C}_{60}$ cage featured the hexagonal facet roughly parallel to the anatase terrace. The optimal surface adsorption site was located above the under-coordinated $\mathrm{Ti}_{5c}$ surface atom, the site identified as most reactive on this surface by earlier studies of small adsorbates \cite{Tilocca:2004uu,He:2009cl}. 

%%% FIG 5. %%%
\begin{figure*}[bhtp]
\centering
\includegraphics[width=15cm]{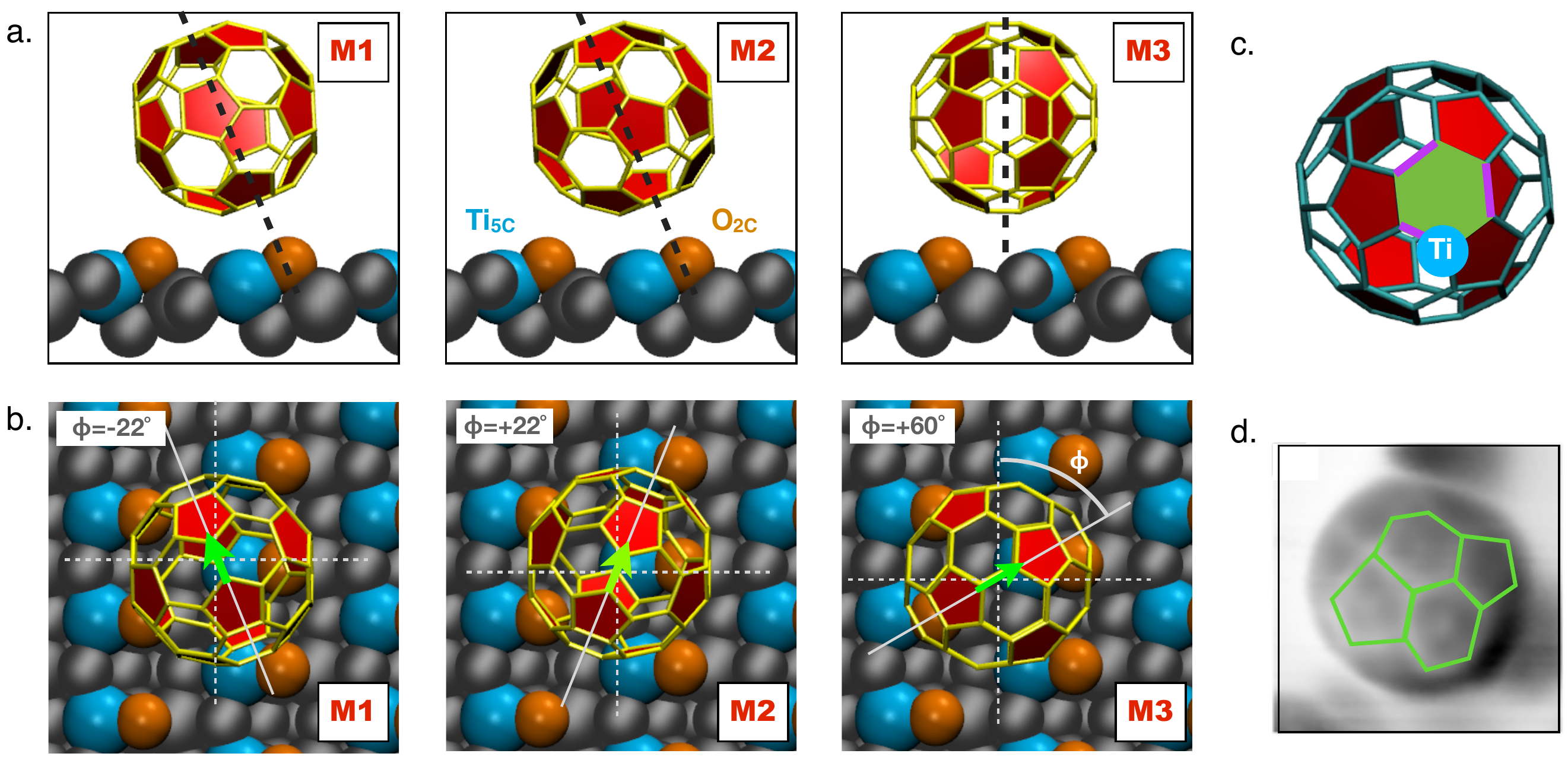}
\caption{Verifying BOSS-predicted structures. a. Side view of the three lowest energy adsorption configurations M1, M2 and M3 obtained by full structural relaxation from BOSS-predicted minima. Pentagonal facets of $\mathrm{C}_{60}$  are coloured in red for visual distinction and the symmetry axis for molecular rotation is indicated by the black dashed line. Reactive under-coordinated atoms on the surface are shown in blue ($\mathrm{Ti}_{5c}$) and orange ($\mathrm{O}_{2c}$) to highlight molecular registry on the surface; b.  Top view of the three lowest energy adsorption configurations M1, M2 and M3. The green arrow illustrates the direction of the typical oval feature observed in all three structures, along the bond between two hexagons. Angle $\phi$ denotes the orientation of the bond with respect to the [010] crystallographic direction; c. Underside of the $\mathrm{C}_{60}$ cage directly above the $\mathrm{Ti}_{5c}$ surface binding site. Molecular binding is facilitated by the hexagonal facet (green) and the nearby $\mathrm{C}_{\mathrm{h}}$-$\mathrm{C}_{\mathrm{h}}$ bond (purple); d. Frequency shift response sub-molecular AFM image of $\mathrm{C}_{60}$ on the (101) surface of $\mathrm{TiO}_2$ anatase, with green lines indicating the top facets. Adapted with permission from Moreno, et. al., Nano Letters, 12, 2257 (2015). Copyright (2015) American Chemical Society.
}
\label{fig5}
\end{figure*}
%%%%%%%%%%%%%%
\textbf{Verifying BOSS-predicted structures.}
The BOSS AES search converged with a global energy minimum of $\mathrm{E}_{\mathrm{BOSS}}$=-1.9eV within the constraint of the structural 'building blocks'. To verify the quality of the prediction, we removed this approximation and allowed all degrees of freedom to relax in DFT. The structure remained the same, with the overall shift in all atomic positions described by a nominal root mean squared distance (RMSD) of 0.19\AA. The resulting global minimum $\mathrm{E}_{\mathrm{GL}}$=-2.0eV (0.1eV below the AI value) and the minimal change in bond lengths (below 0.01\AA) indicated that the 'building block' approximation was appropriate in this case study. 

Next, we compared predicted structures with experimental observations.
In addition to the global minimum, we considered the nearest six unique local minima located by BOSS within a 0.1eV energy window from the 5D global minimum. This allowed us to compare a range of low-energy adsorption configurations with experimental structures, where molecules evaporated onto a hot surface may have acquired similar thermal energy. After seven full structural optimisations, all structures were reduced to one of three different configurations in Fig. \ref{fig5}a. 
%% Justify here!!!

The M1 adsorption geometry was qualitatively identified as the BOSS-predicted 5D global minimum, with M2 as its degenerate mirror image (by 180\degree\ rotation about the axis perpendicular to the anatase terrace). A slight tilt allowed a nearby $\mathrm{C}_{\mathrm{h}}$-$\mathrm{C}_{\mathrm{h}}$ bond to also approach the surface (see \ref{fig5}c.).
The more symmetric M3 configuration in Fig. \ref{fig5}a. was the only local energy minimum found, with an energy of $E_{loc}$=-1.93eV. 
The 5D BOSS search thus led us to non-symmetric low energy configurations stabilised by competing interactions. 
Any symmetric initial guess structure would likely have failed to reach the deeper energy minimum during structure optimisation.

An AFM experimental image with submolecular resolution of $\mathrm{C}_{60}$ on the surface of $\mathrm{TiO}_2$ anatase is presented in Fig. \ref{fig5}d.
For comparison, we considered the top-down view of the three absorption configurations in Fig. \ref{fig5}b.  An elliptical feature with two hexagonal and two pentagonal facets is visible at the top of the molecules. We defined the direction of the feature along the bond separating the two hexagons (the long axis of the ellipse) and computed its orientation with respect to the [010] crystallographic direction to serve as an identification fingerprint. A similar elliptical  feature in the AFM image points to good qualitative agreement between experiment and theory. 
The M1 and M2 molecular structures are topped by a central C atom at the edge of the C-C bond, just like in the experimental image (other BOSS local minima structures were topped by a C-C bond, and we found none topped by a planar facet as in Fig. \ref{fig2}a).
The lack of substrate information made it difficult to conclusively identify the experimental structural fingerprint.

\textbf{Sampling efficiency.} To evaluate the efficiency of BO in structure search, we consider the number of sampled configurations required to converge the global minimum prediction, and later, the AES landscape model. We are not aware of other structure search methods that could provide a comparison. Instead, we compare our method against conventional techniques for determining molecular adsorbate structures: grid-based sampling and human intuition  paired with geometry optimisation. 

BOSS was quick to locate the global minimum in all test cases. 1D  the 2D global minima were identified after 10 and 30 visited configurations respectively. Predictions converged with 150-300 data points in various 3D-4D cases, and 370 in the 5D case. This is a remarkably low computational effort given the vast search space.

In computing the energy landscapes, the number of required data points rose with search dimensionality as well as the complexity of the search (number of minima).  All the preliminary 1D models in Fig. \ref{fig2} required less than 12 data points to converge, at least twice as fast as the grid-based computation of the true energy function with the same resolution. 
In 2D BOSS tests, the $x-y$ landscape was obtained after 45 data points (one minimum), but the more complex $\alpha-\beta$ one required 90 acquisitions (16 minima). The same resolution in the $\alpha-\beta$ AES would require some 500 acquisitions with grid-based methods. 

Grid searches become impractical beyond 3D, whereas BOSS produced good quality AES models also in 3D and 4D simulations (not shown here). These could be sliced in 2D to facilitate the interpretation of the molecule-surface interactions. The many reactive sites of the symmetric $\mathrm{C}_{60}$ cage presented a major challenge for learning the entire AES in 5D, yet BOSS resolved it with only 700 data points. In an intuition-led force minimisation adsorption study, such a computational effort would yield optimised structures from 20-30 different initial guess configurations (assuming that every structure relaxation converges in 20 to 30 single-point DFT calculations). We might choose the best candidate between them, with no possibility of checking if any unknown lower energy structures exist. With AI, 700 data points deliver the optimal configuration across the entire phase space, and additionally, all the local minima and the barriers between them. 

\textbf{Discussion.} We developed an AI-based structure search technique for complex materials that is in line with our ideal methodology described in the Introduction. The BOSS scheme is certainly (i) \emph{efficient} and (ii) \emph{accurate}  in finding the global minimum in 6D (350 DFT evaluations) compared to the traditional structure search strategy. Ultimately, fewer than 100 evaluations would be desirable and further method development (accounting for energy gradients and material symmetries) should considerably speed up the inference. The (iii) \emph{comprehensive} nature of the scheme (global and local minima available) comes at the cost of further computational effort, but the type and the amount of information obtained by inferring the entire energy landscape is not available from other structure search methods. Designing methodology to extract minimum energy paths from N-dimensional energy landscapes would make our scheme even more comprehensive.

Our case study indicates that BOSS is a (iv) \emph{transferable} technique since it inferred both fast and slow varying energy functions by successfully converging parameters on the fly (Fig. 3). Nevertheless, further work on diverse test cases is needed to better characterise method transferability. BOSS  is designed for general degrees of freedom, which facilitates (vi) \emph{flexibility} in workflows with other ML-based structure search techniques. It could be employed for global conformer search of small molecules before these are inserted into the GAtor genetic algorithm scheme for organic crystal structure search \cite{Curtis:2018bu}, or for determining adsorption structures of individual molecules to be employed in registry-based film morphology studies \cite{Obersteiner:2017ei}.

BOSS is certainly (v) \emph{versatile}, since multiple energetic and electronic structure properties are available from each DFT acquisition. Consequently, the inference could be targeted to optimal properties or multi-target objectives instead. It appears straightforward to extend BOSS to (vii) \emph{multi-scale} molecular film simulations, but method performance with increasing dimensionality requires thorough characterisation. Bayesian optimisation scaled better than expected up to 6D (not exponential) on account of periodic kernels employed, and in future work we plan to carry out a quantitative analysis of dimensionality scaling for different tests cases. In our ultimate goal of predicting film formation and morphology we have achieved the first step of having an efficient method for individual molecules on surfaces. We can now build on this to extend BOSS to higher dimensionality (i.e. more than one molecule) or couple it to multi-scale schemes tailored for molecular ensembles.

\textbf{Conclusions.} We proposed a novel structure search scheme that combines a smart AI sampling strategy and a natural "building block" representation with accurate quantum mechanical calculations. As the first step in targeting the structure of large-scale molecular films and organic/inorganic interfaces, we employed it to learn the adsorption structure of a single molecule: $\mathrm{C}_{60}$ on the (101) surface of $\mathrm{TiO}_2$ anatase. 

The BOSS approach facilitated a computationally tractable study of molecular adsorption as a function of key degrees of freedom, molecular registry and orientation. The correct global minimum, verified against fully optimised structures, was located in multi-dimensional phase space with considerable efficiency. Structures based on BOSS-inferred models were in good agreement with high-resolution experimental images of this material. Additional sampling allowed us to compute multi-dimensional AES energy landscapes, with meaningful local minima and energy barriers between them. The resulting chemical insight into the molecule-surface interactions helped us interpret the predicted adsorption structures. Future model refinement could be made more robust by using GP prior belief functions, different GP kernels and by explicitly accounting for material symmetry. 

The 'building block' approach served very well for $\mathrm{C}_{60}$ adsorbed on $\mathrm{TiO}_2$ anatase, and will allow us to readily extend our approach to multi-scale simulations. In short, our BOSS scheme delivers on many fronts in a successful study of molecular surface adsorption and further work will see it applied to more complex configurational studies of surface-supported molecular aggregates and films.

\vspace{0.5cm}
\section{\label{Methods} Methods}

\scriptsize
\textbf{AI software.} BOLFI based on the {\it gpml} package \cite{Rasmussen:2006vz} was implemented in a serial MATLAB code, which was interfaced with the total energy simulation method. 
The knowledge about the PES was encoded in the Gaussian Process (GP), characterised by the GP posterior mean (PES model) and variance functions. The posterior variance supplied a measure of uncertainty on the probabilistic model. We employed a non-isotropic standard periodic GP kernel to account for periodic boundary conditions. Initial sampled data points were selected by a Sobol quasi-random sequence generator, upon which the BO process was initialised. The scheme features only two hyperparameters, which are also learned on-the-fly. The GP model and its hyperparameters were updated every 10 acquisitions until convergence. We analysed the standard deviation on the GP posterior mean: this error remained 0.1eV on average, or 6\% of the energy minimum. We also monitored model quality by noting the convergence of local and global minima, as well as qualitatively checking model cross-sections for the expected symmetries of the atomic model.

\textbf{First-principles calculations.} We performed all configurational sampling with the all-electron DFT code FHI-aims \cite{2009CoPhC.180.2175B}. Simulations were carried out with converged Tier 2 basis sets free of g and h functions, and the PBE exchange-correlation functional \cite{1996PhRvL..77.3865P} augmented with van der Waals correction terms \cite{2009PhRvL.102g3005T}. Relativistic corrections accounted for heavy elements. \emph{Light} grids with $\Gamma$-point reciprocal space sampling was employed to build the PES model. Global minima structures were verified with \emph{tight} grids and a 2$\times$2$\times$1 k-point mesh, which lead to the same geometries, but reduced the adsorption energy by 0.3eV. With the efficient code parallelisation \cite{Marek:2014bh}, a single acquisition calculation on 168 atoms required 10 min on 120 CPUs. The (101) $\mathrm{TiO}_2$ anatase surface slab featured three typical trilayers in a 10.27\AA $\times$11.36\AA$\times$52.77\AA\ periodic unit cell, exposing a 1$\times$3 unit cell surface area \cite{Stetsovych:2015ij}. Molecular adsorption energies converged with three trilayers; the lowest two trilayers were kept fixed during structural optimisations. 

To define the boundaries of BOSS search phase space, we relied on the surface and molecule symmetry and periodicity. Molecular registry search space was limited to the smallest periodically repeating surface unit 10.27\AA$\times$3.78\AA\ and informed by this periodicity. The non-periodic z variable search was conducted 10\AA\ in height from the 1.5\AA\ closest surface approach. 
The high symmetry of the $\mathrm{C}_{60}$ cage was broken by the asymmetric surface features, allowing us to take limited advantage of molecular symmetry. Molecular orientation search was conducted in minimal unique periods of 180\degree\ for $\alpha$ and $\beta$ angles, and 120\degree\ for the $\gamma$ angle, exploiting the symmetry of the $\mathrm{C}_{60}$ cage. The local minimum reference configuration in Fig. \ref{fig2}a, was employed to initialise the BOSS search and set the values for fixed variables when required: (x,y) to (0,0) coordinates in Fig. 4b. (approximately the mid-point between two O2c sites on the surface), z=2.2\AA\ above the surface, and the angles to (0\degree\ , 0\degree\ , 0\degree\ ) as indicated by Fig. 2a.

\textbf{Data availability.} The dataset generated during the study is available as supplementary material.

\textbf{Acknowledgements.} This work was supported by the Academy of Finland through Project Nos. 251748, 284621 and 316601, and also through the European Union's Horizon 2020 research and innovation program under Grant agreement No. 676580 with The Novel Materials Discovery (NOMAD) Laboratory, a European Center of Excellence.  J.C. was funded by the ERC grant no. 742158. Computer time was provided by the Centre for Scientific Computing (CSC, Finland) at the Taito supercomputer.

\textbf{Competing interests.} The authors declare that there are no competing interests.

\textbf{Author contribution.} P.R., J.C., M.U.G. and M.T. designed the research. M.U.G. contributed BO routines, M.U.G. and M.T. produced the BOSS code. M.T. performed the research, analyzed the data and wrote the manuscript. All the authors read the manuscript, commented the contents and agreed with the publication of the results. J.C.and P.R. contributed equally to this work.

%\bibliography{Todorovic_et.al._npjComputMater}


\end{document}